\documentclass[letterpaper,twocolumn,10pt]{article}
\usepackage{usenix,epsfig,endnotes}
\usepackage{textcomp}
\usepackage[latin1]{inputenc}
\usepackage{amssymb}
\usepackage{listings}
\usepackage{epsf,amsmath}
\usepackage{subfig}
\usepackage{graphics}
\usepackage{multirow} 
\usepackage{color}
\definecolor{orange}{rgb}{1,0.5,0}

\newcommand{\ie}{{i.\,e.}}

\renewcommand{\epsilon}{\varepsilon}

\begin{document}

\date{}

\title{\Large \bf Highly-Concurrent Doubly-Linked Lists \vspace{-6mm}}

\author{
{\rm Nitin Garg}\\
\and
\hspace{5mm}
{\rm Ed Zhu}\\
\and
{\rm Fabiano C. Botelho} \\[4mm]
\hspace{-75mm}
Data Domain, an EMC$^2$ Company,
Santa Clara, CA, USA \\
\hspace{-75mm}
\texttt{\{nitin.garg, ed.zhu, fabiano.botelho\}@emc.com}
\vspace{-40mm}
} 
\maketitle

\thispagestyle{empty}

\enlargethispage{\baselineskip}

\subsection*{Abstract \vspace{-5mm}}
As file systems are increasingly being deployed on ever larger systems with many cores and multi-gigabytes of memory, scaling the internal data structures of file systems has taken greater importance and urgency. A doubly-linked list is a simple and very commonly used data structure in file systems but it is not very friendly to multi-threaded use. While special cases of lists, such as queues and stacks, have lock-free versions that scale reasonably well, the general form of a doubly-linked list offers no such solution. Using a mutex to serialize all operations remains the de-facto method of maintaining a doubly linked list. This severely limits the scalability of the list and developers must resort to ad-hoc workarounds that involve using multiple smaller lists (with individual locks) and deal with the resulting complexity of the system.

In this paper, we present an approach to building highly concurrent data structures, with special focus on the implementation of highly concurrent doubly-linked lists. Dubbed ``advanced doubly-linked list'' or  ``adlist'' for short, our list allows iteration in any direction, and insert/delete operations over non-overlapping nodes to execute in parallel. Operations with common nodes get serialized so as to always present a locally consistent view to the callers. An adlist node needs an additional 8 bytes of space for keeping synchronization information -- this is only possible due to the use of light-weight synchronization primitives called lwlocks which are optimized for small memory footprint. A read-write lwlock occupies 4 bytes, a mutex occupies 4 bytes, and a condition variable occupies 4 bytes. The corresponding primitives of the popular pthread library occupy 56, 40 and 48 bytes respectively on the x86-64 platform. The Data Domain File System makes extensive use of adlists which has allowed for significant scaling of the system without sacrificing simplicity.

\sloppy
\lstset{
  language=C,
  basicstyle=\fontsize{8}{8}\selectfont,
   aboveskip=0mm,
   belowskip=0mm,
   mathescape=true,
   escapechar=@,
   showstringspaces=false,
   columns=fixed,
   basewidth=0.515em,
   frame=none,
   framesep=1mm,
}

\section{Introduction\vspace{-5mm}}
\label{introduction}

The simplest data structures often get used the most. This is particularly true for linked lists which are used in a wide variety of ways in most file systems. Linked lists can be used to hold incoming IO requests, for simple LRUs caches, for maintaining work queues or other FIFO data structures. While simple and useful, a generic doubly-linked list does not scale well with respect to concurrent access. 

As hardware platforms support increasing number of cores, having basic data structures be as concurrent as possible is now a prerequisite~\cite{s11}. There have been good strides in building lock-free or highly-concurrent versions of simpler variants of lists, such as LIFO and FIFO queues~\cite{K03,ls04}, singly-linked lists~\cite{Harris01apragmatic}, or double-ended queues~\cite{Sundell08lock-freeand}, and many more variants~\cite{Attiya06built-incoloring}. There are a number of published lock-free approaches both for any data structure in general~\cite{Detlefs01lockfreereference,Herlihy93transactionalmemory} and for generic doubly-linked lists, in particular~\cite{Sundell08lock-freeand, paulmartin}. All lock-free approaches, however, either do not support true iterators, or implicitly rely on the system having a strong garbage collection: otherwise, an iterator having read the next pointer from a node will have no guarantee that the next node will be valid when the pointer is dereferenced. Some lock-free approaches are also not entirely practical owing to their reliance on hardware features not commonly available or the complexity of the approaches. Due to these limitations, we do not presently consider these lock-free approaches viable for our needs in building file systems.

The default approach, therefore, for protecting a generic doubly-linked list in a multi-threaded environment with no special system support remains the use of a single lock to protect the entire list during any supported operation. Attempts to increase concurrent access by using localized locks usually runs afoul of one or more of the following issues: (i) the memory overhead is too high when using per node locks, (ii) the need to enforce ``lock ordering'' restricts supportable API or (iii) the supported API is prone to spinning behavior and restricts member nodes to preallocated memory.

In this paper, we present our approach to building highly-concurrent doubly-linked lists. While we focus primarily on the list data structure, the general ideas presented are applicable to other data structures as well. We call our highly-concurrent doubly-linked list an ``advanced doubly-linked list'' or ``adlist'' for short. An adlist supports iteration (forward or backward), insert before/after any node (or at head/tail of list) and delete of any specified node (or dequeue/pop from list). It also places no restrictions on the number of nodes a list can have or where the memory for the nodes comes from. The nodes can be dynamically allocated and returned to the system memory once removed from the list. In short, it supports most of the operations that one would expect to perform on a generic doubly-linked list. Operations that involve disjoint sets of nodes proceed in parallel. Operations with common nodes get serialized so as to always present a locally consistent view to the callers. The operations are not lock-free but threads do not spin when contending on a node.

The low memory overhead per node is possible due to the use of ``lwlock''~\cite{2011arXiv1109.2638G} primitives. The ``asynchronous'' or ``deferred blocking'' mode of lwlocks also provide the crucial framework for dealing with lock ordering issues. The synchronization structures add an overhead of 8 bytes per node. The adlist was developed for and is used extensively in the Data Domain File System~\cite{zlp08}. We have found it to lower contention and improve scalability of our system significantly without sacrificing simplicity of code.

The rest of the paper is organized as follows: Section~\ref{lwlock_intro} summarizes the basic idea behind lwlocks and asynchronous-locking as used in adlists. Section~\ref{node_internal} describes the internal structure of the synchronization primitive of an adlist node. Section~\ref{adl_api} walks through the APIs supported by an adlist highlighting how we maintain correctness. In section~\ref{experience}, we discuss our observations and experience with using adlists as well as touch upon possible extensions. In section~\ref{performance} we present experimental evaluation of adlist performance. Finally, in section~\ref{conclusion} we present our conclusions.

\section{Light-weight locks}
\label{lwlock_intro}

Light-weight locks~\cite{2011arXiv1109.2638G} or lwlocks are synchronization primitives that are optimized for small memory footprint while maintaining performance comparable to pthread library's synchronization primitives in low contention cases. A read-write lwlock occupies 4 bytes, a mutex occupies 4 bytes (2 if deadlock detection is not required), and a condition variable occupies 4 bytes. The corresponding primitives of the popular pthread library occupy 56 bytes, 40 bytes and 48 bytes respectively on the x86-64 platform. 

The core idea behind lwlocks is the observation that while a thread could block on different locks or wait on many different condition variables in its lifetime, it can block on only one lock or condition variable at any given point. With lwlocks, whenever a thread has to block, it uses a ``waiter'' structure to do so.
Each thread has its own waiter structure and can access it by invoking the \texttt{tls\_get\_waiter} function (which returns the pointer to the waiter kept in the thread local storage).

Figure~\ref{tls} presents the definition of a waiter structure. For compact representation, the maximum number of waiter structures is limited to be less than $2^{16}$ so that each structure can also be uniquely referred to by a 16-bit number. The value $2^{16} - 1$ is used to represent the \texttt{NULL} waiter structure and denote it by \texttt{NULLID}. The limit on number of waiters, and hence on the number of threads, is large enough for most applications and certainly so for use in adlists.
\begin{figure}
\begin{center}
\begin{lstlisting}
	  @{\em // Assume that each bool\_t takes 4 bytes, each}@
	  @{\em // pthread\_mutex\_t takes 40 bytes, each pthread\_cond\_t}@ 
	  @{\em // takes 48 bytes, and each function pointer takes 8 bytes.}@ 
	  @{\bf interface}@ event_t {
	    void signal();
	    void wait();
	    bool_t poll();
	  } @{\em // 24 bytes}@

	  @{\bf interface}@ domain_t {
	    waiter_t alloc_waiter();
	    void free_waiter(waiter_t waiter);
	    waiter_t get_waiter();
	    waiter_t id2waiter(uint16_t id);
	  } @{\em // 32 bytes}@

	  struct waiter_t {
	    event_t  event;
	    domain_t domain;
	    bool_t signal_pending;
	    bool_t waiter_waiting;
	    pthread_mutex_t mutex; 
	    pthread_cond_t  cond; 
	    uint64_t app_data;
	    uint16_t id;
	    uint16_t next;
	    uint16_t prev;
	  } @{\em // 166 bytes}@
\end{lstlisting}
\end{center}
\caption{Definition of a waiter structure.}
\label{tls}
\end{figure}

A waiter structure is assigned to a thread the first time the thread accesses it (via \texttt{tls\_get\_waiter}) and the structure is returned to the pool of free waiter structures when the thread exits to be re-used by a later thread. The waiter structure is the key piece that enables the compact nature of lwlocks. It can also be used to create other custom compact lock-like data structures, a feature that is used in adlist (see section~\ref{node_internal}).

{\bf Waiter's Event.}
A generic event interface underlies the actual mechanics that are used by a thread when it blocks or unblocks on a lock. The two main operations defined for an event are: (i) \texttt{wait}, which is called to wait for the event to trigger; and (ii) \texttt{signal}, which informs a waiter of an event getting triggered. A waiter structure uses one pthread mutex and one condition variable to implement both operations. The operation \texttt{wait} blocks the thread on the condition variable until a signal arrives. The operation \texttt{signal} wakes up the blocked thread. Like semaphores, the implementation ensures that a signal on an event cannot be lost, \ie, a \texttt{signal} can be invoked before the matching \texttt{wait} is and the \texttt{wait} will find the pending signal. Unlike a semaphore, however, the operations \texttt{wait} and \texttt{signal} are always called in pairs. There is also a third operation called \texttt{poll}. It can be used to check if a signal is already pending.

{\bf Forming Lists or Stacks of Waiters.}
Each waiter structure records its own id. It also has space for previous and next id values which can be used to form stacks or lists of waiters. Such a list (or stack) of waiters can be identified purely by the id of the first element of the list, \ie, it can be represented by a 16-bit value. To go to the next (previous) waiter, we convert the current id to the corresponding waiter structure and look at the next (previous) id field in it. 

{\bf Locking Data.}
The final important piece of a waiter structure is the space it provides that can be used by the abstractions built on top for their own purpose. The waiter itself does not interpret it in any way. For instance, read-write lwlocks use this space to record the type of locking operation that the thread was performing when it blocked: whether it was taking a read or write lock. This space amounts to 8 bytes and is referred to as \texttt{app\_data}.

\subsection{Reader-writer lwlock}
\label{rw_lwlock}
Adlists use a reader-writer lwlock in each adlist node and it accounts for 4 out of the 8 bytes of synchronization space. Henceforth, we refer to a reader-write lwlock as simply a lwlock. An lwlock by default is ``fair'': the lock is acquired in FIFO order by the threads blocked on it and a thread that wants to acquire an lwlock will block if there are already other threads waiting to acquire the lock. Pthread locks are not fair in this sense, and although it is possible to build lwlocks to mimic the same behavior, we have found fairness to be the better option in general. Adlists, in particular, rely on the fairness characteristics of lwlocks.

An lwlock uses 2 bytes to keep a queue of waiter structures of the threads that have requested the lock which is currently held. This queue is aptly called a \texttt{waitq}. The \texttt{waitq} is maintained as a ``reverse list'' as that allows insertion of a new waiter in a single hardware supported compare-and-swap (CAS) instruction. The \texttt{next} field of a waiter structure holds the id of the waiter structure in front of it. The oldest waiter's \texttt{next} field holds \texttt{NULLID}. Another important feature of fair lwlocks is that lock ownership is transferred by the owner during an \texttt{unlock}. The unlocking thread has to walk the \texttt{waitq} to find the waiter(s) to \texttt{signal}. It removes the selected waiter(s) from the \texttt{waitq} before calling \texttt{signal}. At any point there can be only one thread performing the transfer on a lock and hence the walk and \texttt{waitq} modification is safe to perform. Since the unlocking thread does the work of transferring the lock state and ownership, the thread corresponding to a waiter can and should assume it has the lock once \texttt{signal} is called on the waiter.

Of the remaining 16 bits of an lwlock, 14 bits are used for the count of read locks granted, 1 bit is used to indicate a write lock, and the final 1 bit is used to indicate whether the lock is read-biased or not. A read-biased lock is unfair towards writers in the sense that a thread that needs a read lock will acquire it without any regard to waiting writers if the lock is already held by other readers. This behavior is similar to that of pthread read-write lock and is essential for applications where a thread can recursively acquire the same lock as a reader. Without the read-biased behavior, a deadlock can result if a writer arrives in between two read lock acquisitions: the second read lock attempt will wait for the writer which is waiting for the first read lock to be released. Applications that do not have recursive read locking do not need the read-biased behavior but may choose to use it for throughput reasons.\footnote{The 14-bit reader count limits the maximum number of readers per lock to $2^{14}$, a limit that we have found to be sufficient in practice for adlists. The limit can be raised by having the API explicitly flag read-bias behavior, so the bias bit does not have to be in the lwlock or restricting the maximum concurrency, thereby freeing bits from the \texttt{waitq} or by slightly increasing the size of the lock.} Adlists do not use read-biased lwlocks and we do not discuss them further in this paper.

Figure~\ref{rwlock_algos} outlines the algorithms for the two main operations: (i) \texttt{lock}, and (ii) \texttt{unlock}.
To acquire a lock, a thread uses the CAS instruction to either take ownership of the lock or add its own waiter structure to the lock's \texttt{waitq}. If the lock is acquired, nothing more needs to be done. If it is not acquired, then the thread
has to \texttt{wait} for the \texttt{signal} to arrive. We note that the lock operation is based on the novel asynchronous locking functionality of lwlocks, which is described in the following section. The unlock operation has to pick the oldest set of waiters that it can signal: either a single writer or a set of contiguous readers, and change the lock state accordingly. The algorithm outline in figure~\ref{rwlock_algos} is only at a high level for the contention case on non-read-biased lwlock. The non-contented case is simple to derive. We encourage interested readers to get all the details about the workings of all the lwlock primitives in~\cite{2011arXiv1109.2638G}.

\begin{figure}
\vspace{-3mm}
\begin{center}
\begin{lstlisting}
struct lwlock_t {
  uint1_t rd_bias;  
  uint1_t wlocked;  
  uint14_t readers; 
  uint16_t waitq; 
}
bool_t @async\_lock(lwlock\_t $\ell$, bool\_t exclusive)@ {
  @$w$@ = tls_get_waiter();
  do {
    @$o$@ = @$n$@ = @$\ell$@;
    if (!exclusive && !@$o$@.wlocked && 
       (@$o$@.waitq @==@ @NULLID@ || 
        @$o$@.rd_bias))
      @$n$@.readers++;
    else if (@exclusive@ && 
            !(@$n$@.wlocked || @$n$@.readers > 0))
      @$n$@.wlocked = 1;
    else { @{\em // Need to block}@
      @$w$@.app_data = exclusive;
      @$w$@.next = @$o$@.waitq;
      @$n$@.waitq = @$w$@.id;
    }
  } while (!CAS(@$\ell$@, @$o$@, @$n$@));
  if @($n$.waitq == $w$.id) return FALSE;@
  else return TRUE;
}
void @lock(lwlock\_t $\ell$, bool\_t exclusive)@ {
  if @(!async\_lock($\ell$, exclusive)) $w$.event.wait();@
  return;
}
uint16_t waitq_size(uint16_t @$wid$@) {
  uint16_t count = 0;
  while (@$wid$@ != @NULLID@) {
    @$wid$@ = id2waiter(@$wid$@).next;
    count++;
  }
  return count;
}
void @unlock\_fair(lwlock\_t $\ell$)@ {
  do {
    @$o$@ = @$n$@ = @$\ell$@;
    if (@$n$@.wlocked == 1) @$n$@.wlocked = 0;
    else @$n$@.readers--;
    if(!(@$n$@.wlocked || @$n$@.readers > 0)) { 
      @$(pw, wtw)$ = find\_oldest\_set\_of\_waiters($n$)@;
      if (@$pw$@ == @NULL@) @$n$@.waitq = @NULLID@;
      if (@$wtw$@.app_data != exclusive) { 
        @$n$@.readers = waitq_size(@$wtw$.id@);
      } else @{\em // single writer picked}@
         @$n$@.wlocked = 1;
    }    
  } while (!CAS(@$\ell$@, @$o$@, @$n$@));
  if (@$pw$@ != @NULL@) @$pw$@.next = @NULLID@;
  wake_up_waiters(@$wtw$@);
}
\end{lstlisting}
\end{center}
\vspace{-4mm}
\caption{Operations to lock and unlock a lwlock. 
The \texttt{lock} operation takes a boolean as input to indicate whether an exclusive lock 
is requested. The old and new values passed in to CAS are denoted 
by $o$ and $n$, respectively. The caller's thread local waiter structure is denoted by $w$.
We use $wtw$ and $pw$ to denote the waiter to wake up, and the 
waiter before $wtw$ in the \texttt{waitq} respectively.
}
\vspace{-4mm}
\label{rwlock_algos}
\end{figure}

\subsection{Asynchronous locking}
\label{async_lock}

A key observation about lwlock's locking behavior is that once a thread has put itself on the \texttt{waitq} of a lwlock, it is guaranteed to have the lock transferred to it. The thread does not have to call \texttt{wait} right away. It has to call \texttt{wait} eventually but it can perform other actions before calling \texttt{wait}. Upon returning from \texttt{wait}, a thread can assume that it holds the lock in the mode it requested. This observation is reflected in the pseudo-code in figure~\ref{rwlock_algos} and forms the basis of asynchronous locking.

The asynchronous locking functionality of lwlocks is implemented by the \texttt{async\_lock} function. If the lock could not be acquired, it does not call \texttt{wait} itself but returns a boolean so the caller knows if it should. 
Calling \texttt{async\_lock} is similar to calling a \texttt{trylock} except the caller is then guaranteed to have the lock assigned to it at some future point. The caller must eventually call \texttt{wait} to acknowledge the ownership and once done, \texttt{unlock} the lock. Since each thread has only 1 waiter structure, once a call to \texttt{async\_lock} returns FALSE, the thread must not perform any action that would result in the structure being used again before it has called \texttt{wait}. Failure to adhere to this rule will most likely lead to one or more of system hang, corruption or crash.

Asynchronous locking is the key enabler to work around the constraints that lock-ordering imposes. An adlist allows concurrent appends, dequeues, inserts (before or after any member), deletes and iterators (in either direction). Some of these operations need to acquire locks in opposite order of other operations. To avoid deadlocks, a canonical order is picked (follow the \texttt{next} pointer) and operations that need to acquire locks in the opposite direction (along \texttt{prev} pointer) use asynchronous locking. 

\subsubsection*{Using asynchronous locking for adlists}
\label{why_async_lock}

The following example illustrates how asynchronous locking is used and why it is essential. Suppose the canonical order for nodes A \& B is A, then B. A thread holds a lock on B already and needs to lock A. It will make an asynchronous lock call for A. If the thread is unable to get the lock, it is on A's \texttt{waitq}, and it releases the lock on B. It then waits for the lock on A to be granted and then reacquires the lock on B (which is in canonical order). In the above sequence, the thread always either holds a lock (on A or B) or is in the \texttt{waitq} of a lock (on A). Other guarantees in adlist implementation ensure that in this case A and B will remain valid and hence there will be no illegal access. Achieving this without asynchronous locking is not possible. Using \texttt{trylock} on A and upon failure, releasing B then locking A leaves a window open between release of B and locking of A where neither node is in any way aware of the thread. One or both nodes could go away in that window and the thread would end up performing an illegal access.

\section{Internals of an adlist node}
\label{node_internal}

Figure~\ref{adl_node} shows the internal fields of an adlist node. The \texttt{prev} and \texttt{next} pointer take up 16 bytes, the lwlock takes up 4 bytes and the remaining 4 bytes are taken up by an \texttt{adl\_refcnt\_t} structure. The \texttt{adl\_refcnt\_t} structure is a custom synchronization primitive that is built using the same general principals as other lwlock primitives. The \texttt{lock} field protects the \texttt{prev} and \texttt{next} pointers. The lock must be acquired in shared mode for reading either of the pointers and in exclusive mode for changing either of the pointers. We currently use a single lock to protect both the pointers but it is not necessary to do so. Using separate locks for the 2 pointers would allow greater concurrency at the expense of the additional overhead. For now a single lock has sufficed for our needs. The adlist structure itself contains 2 fixed nodes, the \texttt{head} and \texttt{tail} nodes between which all user added nodes are kept.

\begin{figure}
\begin{center}
\begin{lstlisting}
    struct adl_refcnt_t {
      uint1_t mask; 
      uint15_t pincount; 
      uint16_t waitq; 
    } @{\em // 4 bytes}@
    struct adl_node_t {
      adl_node_t  *next;
      adl_node_t  *prev;
      lwlock_t    lock;
      adl_refcnt_t refcnt;
    } @{\em // 24 bytes}@
\end{lstlisting}
\end{center}
\caption{Internal structure of an adlist node.}
\label{adl_node}
\end{figure}

\subsection{adl\_refcnt\_t}
\label{adlrefcnt}

The \texttt{adl\_refcnt\_t} structure provides synchronization between read-only and modifying operations on an adlist. As can be seen in figure~\ref{adl_node}, its internal structure is somewhat analogous to that of an lwlock. Like a lwlock, the structure is also manipulated using CAS instructions. The behavior, however, is quite different. We now go over the details of each of the fields.

\subsubsection{Mask bit}
\label{refcnt_mask}
The mask bit controls the visibility of a node in the adlist. If the mask bit is set the node is considered to be invisible. Iterators will skip over this node and not return it to the upper layers. Attempt to set the \texttt{mask} when it is already set returns an error to the caller. Hence the act of setting or unsetting the \texttt{mask} acts as a serialization point between an iteration and delete operation, between two delete operations, and between an iteration and insert operation.

A node delete operation starts by first setting the \texttt{mask} bit. If two threads attempt to delete the same node, then only the thread that sets the \texttt{mask} will continue with the delete while the other will return failure to the caller. The moment when the \texttt{mask} is set is considered to be the moment that the node was removed from the list. Any iterators reaching the node after that will skip it as if it doesn't exist effectively serializing the operation as delete first, then iterate. If an iterator had already returned the node earlier, then the serialization is iterate first, then delete. Note that setting the \texttt{mask} only marks the point in time when the delete took place. The actual deletion of the node which involves updating the pointers of the neighboring nodes still needs to be done and it has to co-ordinate with any other competing operations occurring in the same neighborhood. Also note that once a node has been masked for delete, the caller must eventually go through with the delete. There is no option to abandon the intent to delete by clearing the \texttt{mask} bit. This restriction is mainly to keep things simple and we haven't found a compelling use case to support the abandon option.

When a node is inserted into the list, its \texttt{mask} bit is set until all the pointers are set up. Only then is the \texttt{mask} bit cleared. The moment of clearing the \texttt{mask} bit is considered to be the moment that the insert actually occurred and is the serialization point between an insert and an iteration operation.

\subsubsection{Pin count}
\label{refcnt_pin}
The value in the \texttt{pincount} field determines whether a node can be removed from the adlist or not. A non-zero value means that the node cannot be removed. The \texttt{adl\_node\_pin} API can be called to increment the \texttt{pincount} and \texttt{adl\_node\_unpin} can be called to decrement it. The unpin call will always succeed unless the \texttt{pincount} is already zero (which will cause a crash). The \texttt{adl\_node\_pin} only succeeds if the \texttt{mask} bit is not set, otherwise it returns FALSE. Any caller to \texttt{adl\_node\_pin} must be able to handle the failure to pin. There is an internal function, \texttt{adl\_node\_force\_pin}, that can increment the \texttt{pincount} even when the \texttt{mask} is already set if the existing value is already non-zero. This function is not exposed outside the adlist library and is used internally to give priority to iterators over a delete operation.

Iterators pin the node they are on before returning it to the caller. The \texttt{adl\_iter\_next} API returns the ``next'' node based on the direction the iterator was set up with. There are two possible directions: (i) {\it forward} where the \texttt{next} field of a node is followed; and (ii) {\it backward} where the \texttt{prev} field is followed. Maintaining a pin on the returned node ensures that the iterators foothold (the node) will stay in the adlist until the iterator is invoked again to get the next node. Invoking \texttt{adl\_iter\_next} also implicitly releases the pin on the node returned earlier.

\subsubsection{Waitq}
\label{refcnt_waitq}
The \texttt{waitq} is used to co-ordinate between a delete operation that is trying to remove the node from the list and any iterators that are moving backwards and need to pass over the node.

Once a delete operation has set the \texttt{mask} bit, it cannot start on the process of actually removing the node unless the \texttt{pincount} on the node reaches zero. The deleting thread places its waiter's id in the \texttt{waitq} field when it is waiting for the \texttt{pincount} to reach zero. The last thread to call \texttt{adl\_node\_unpin} invokes the \texttt{signal} call of the delete thread's \texttt{waiter}.

An iterator uses the \texttt{waitq} when it has to wait for a delete operation. Let's say an iterator that is going backwards is on node A. Let's say the node before A is B and C is the one before B. Also, B is already masked while C is not masked. The iterator therefore has to skip over B and return C when \texttt{adl\_iter\_next} is called. It also needs to drop the pin it has on A and acquire one on C before returning it. To skip over B, the iterator has to first temporarily pin it. Only then can the iterator safely assume that B will not be removed while it tries to acquire the \texttt{lock} to read the B's \texttt{prev} pointer. Since B is already masked, the iterator tries \texttt{adl\_node\_force\_pin} call to increment the \texttt{pincount}. If the call succeeds, the iterator can drop the the lock and pin on A and restart the operation as if it was always on B and is going to C. If the call fails, then the delete operation had already been signaled to proceed and the iterator must wait for the delete to finish. It does so by adding its waiter to the \texttt{waitq} of B before dropping the lock on A. When the delete operation has finished, it will signal all the waiters on the \texttt{waitq} of B. At this point, B is no longer on the list and A's \texttt{prev} points to C. The iterator still has its pin on A. Hence, it can basically restart the \texttt{adl\_iter\_next} operation.

Note that the use of \texttt{adl\_node\_force\_pin} call implicitly gives priority to iterators over the delete operation. This behavior is not necessary but the rationale is that the forced pin is expected to be released relatively quickly as the iterator tries to skip over the masked node. It does expose the delete operations to the possibility of being starved but we have found it to be satisfactory in practice. Systems where this would be unacceptable can simply make the iterators wait if the node is already masked.

\section{Adlist APIs}
\label{adl_api}

We have already seen hints of the APIs supported by adlists. We now present them systematically and prove that the APIs work reliably when there is contention. The API supported by adlist can be broadly divided into 3 categories. (i) Operations that add nodes to the list, (ii) Operations that remove nodes from the list and (iii) operations that iterate over the list. Table~\ref{api_table} shows the available APIs in each category. APIs that are really convenience wrappers around other calls are marked with an asterisk. Aside from these, there are also the aforementioned \texttt{adl\_node\_pin} and \texttt{adl\_node\_unpin} calls to manage a node's removability in the adlist.

\begin{table}
{\scriptsize
\begin{tabular}{|c|c|c|}
\hline 
{\bf Category} & {\bf API} & {\bf Comment}\\
\hline
\multirow{7}{*}{iterate} & adl\_next & -- \\
\cline{2-3}
                        & adl\_prev &  -- \\   
\cline{2-3}

                        & adl\_first* & -- \\
\cline{2-3}

                        & adl\_last* & -- \\
\cline{2-3}

                        & adl\_iter\_init* & specify direction \\
\cline{2-3}

                        & adl\_iter\_next* & -- \\
\cline{2-3}

                        & \multirow{2}{*}{adl\_iter\_destroy*} & releases pin on \\                                                                       
                        &                                      & current node, if any \\                                                                       
\hline
\multirow{4}{*}{insert} & adl\_insert\_after & -- \\
\cline{2-3}
                        & adl\_insert\_before & -- \\   
\cline{2-3}

                        & adl\_insert\_at\_front* & --\\   
\cline{2-3}

                        & adl\_append\_at\_end* & --\\   
\hline
\multirow{7}{*}{delete} & adl\_node\_remove\_start & mask node \\
\cline{2-3}
                        & adl\_node\_remove\_waitonpincount & wait to start delete \\   
\cline{2-3}

                        & adl\_node\_remove\_do & update pointers \\   
\cline{2-3}

                        & \multirow{2}{*}{adl\_node\_delete*} & wrapper around \\   
                        &                                     & above three \\   
\cline{2-3}

                        & adl\_pop* & remove first node \\
\cline{2-3}

                        & adl\_dequeue* & remove last node \\
\cline{2-3}

                        & \multirow{2}{*}{adl\_iter\_pop*} & remove current\\                       
                        &                                  & iter node \\                       
\hline
\end{tabular}
}
\caption{Supportable APIs of adlists.} 
\label{api_table} 
\end{table}

Regardless of the operation type of an API call, they all follow some basic rules. All APIs that take a node as an input need a guarantee that the node is on the list. This is usually achieved by having the caller take a pin on the node, either directly or implicitly due to an earlier action, although having the node masked would also work. The only exception, naturally, is the node being added in insert operations which is initialized internally and added to the list. Functions that return a node as their result always return the node in pinned state. Insert operations also leave the new node in pinned state. It is the callers responsibility to drop the pin once it is no longer needed. Convenience wrappers like iterators drop the pin internally on nodes that no longer need to be pinned.

The agreed upon ``canonical order'' of locks in the list is along the \texttt{next} pointer of the nodes. Any lwlocks acquired in the canonical order can be acquired in blocking mode. Any locks acquired in the reverse order must be done
using asynchronous-locking with proper handling of cases where the asynchronous-lock doesn't succeed right away. In other words, if a node B can be reached from node A by following one or more \texttt{next} pointers, then the lwlock of node B can be acquired in blocking mode while holding the lwlock of A. To acquire the lwlock of A while holding the lwlock of B, we must use the \texttt{async\_lock} call. The specific details of how to handle the case when \texttt{async\_lock} returns FALSE varies slightly from API to API but the general essence is to drop the lwlock on B, then wait for the lock on A to be granted. Once acquired, the lock on B can be re-acquired in blocking mode. The natural questions to ask here are how do we ensure that (i) the node B remains valid when we are ready to re-acquire the lock and (ii) that the node B remains in a position after node A so that a blocking lock on node B won't violate the canonical order. We hope to convince the readers in the next portion that we do indeed maintain these properties.

All the API calls are orchestrated carefully so that the above requirements are always honored regardless of what else happens. There are multiple means by which one can ensure that a node stays on the list: Having a pin on the node prevents it from being removed; Setting the \texttt{mask} bit ensure no other thread will try to remove the node; taking the lwlock, in either shared or exclusive mode, of a node or either of it's neighbors ensures that the node cannot be removed as the relevant pointers cannot be updated without acquiring these locks. Starting with the fact that an input node is either pinned or masked, each API call uses one or more of these guarantees in a careful dance with all the other ongoing contending operations to ensure correctness when accessing any nodes required for the call.

\subsection{API Implementation}
\label{api_impl}

Before we go into the contention cases to illustrate how correctness is maintained, let us look at the general structure of some selected core API operations. We will only look at operations that need to take locks in non-canonical order. Operations, such as \texttt{adl\_next} and \texttt{adl\_insert\_after}, that only take locks in canonical order are relatively straight-forward.

{\bf adl\_prev.}
The \texttt{adl\_prev} call takes a node as an input and returns the first node before it which is not masked. The input node is guaranteed by the caller to remain on the list either because it is pinned by the caller or was masked for delete by the caller. When there is no contention, the operation will proceed as follows: (i) Acquire lock of input node in shared mode. (ii) Read the \texttt{prev} pointer. (iii) Call \texttt{adl\_node\_pin} to pin the previous node. (iv) Release lock of input node. (v) Return previous node.

{\bf adl\_insert\_before.}
The \texttt{adl\_insert\_before} call takes an existing node and a new node as input and inserts the new node before the existing node. The existing node must be pinned or masked by the caller. In the no contention case, the operation proceeds as: (i) initialize the \texttt{lock} and \texttt{refcnt} of the new node. The \texttt{refcnt} is initialized as masked and pinned. (ii) Take exclusive lock on input node. (iii) Read \texttt{prev} pointer of input node. (iv) Use \texttt{async\_lock} to take exclusive lock on previous node. (v) Setup the needed pointers to add new node in between input and previous nodes. (vi) Unmask new node to make it visible. It is left pinned. (vii) Release locks on previous and input nodes and then return.

{\bf adl\_node\_delete.}
The \texttt{adl\_node\_delete} operation deletes the input node from the list. Unlike the other APIs above, the input node must be pinned, not masked, by the caller beforehand. It is actually a wrapper around the 3 step node removal process. Step 1 is the call to \texttt{adl\_node\_remove\_start} which tries to set the \texttt{mask} bit of the node. The function can return one of three possible outcomes: (i) The \texttt{mask} could not be set. Return FALSE to caller. (ii) \texttt{mask} is set and \texttt{pincount} was also decremented to 0. Go directly to step 3. (iii) \texttt{mask} is set but the \texttt{pincount} is not 0. The pin held by the caller is not internally dropped in this case as was done in case (ii). The call needs to go through step 2.

Step 2 is the \texttt{adl\_node\_remove\_waitonpincount} function which drops the pin held by the caller and if the \texttt{pincount} is still non-zero, puts the caller's waiter in the \texttt{waitq}. This is done in a single CAS instruction. If the waiter was queued, it then calls the \texttt{wait} function on its waiter. Upon waking up from the waiter, the \texttt{pincount} has reached 0, and the delete can proceed to step 3.

Step 3 is the \texttt{adl\_node\_remove\_do} function which does the work of updating the pointers to remove the node from the list. In the no contention case, the operation proceeds as follows: (i) Take exclusive lock on node. (ii) Read the \texttt{prev} pointer and take exclusive \texttt{async\_lock} on the previous node. (iii) Read the \texttt{next} pointer and take exclusive lock on the next node. (iv) Update the relevant pointers and release the locks on the previous and next nodes. The node is now off the list and cannot be reached by anyone via the list. (v) ``Clear'' the deleted node which involves waking up any thread that was waiting either on the \texttt{lock} or the \texttt{waitq} of the node. The iterators waiting on the \texttt{waitq} are simply signaled. The threads waiting on the lwlock are expecting to have the lock transferred to them. To wake them up properly and wait for them to clear out, the function calls \texttt{async\_lock} on the node's lwlock that it holds itself already. This call will always result in the thread's waiter getting queued on the lwlock. The lwlock is then released which transfers it to the first waiting thread. As the waiting threads get the lock one by one, each realizes the node is no longer of interest and passes the lwlock to the next waiting thread. The lwlock eventually comes back to the thread doing the delete at which point it can be sure that there are no threads in any way trying to access the node. The delete operation is now complete and the node can be disposed off by the caller in any way it sees fit.

\subsection{Dealing with Contention}
\label{api_contention}

We now look at how the various APIs handle contention. The simplest case is the iterator API which we discuss first. Looking at the \texttt{adl\_prev} implementation, the only problem case contention it can run into is in step (iii) when the previous node is already masked. The outline of how we handle this was covered in section~\ref{refcnt_waitq} on the \texttt{waitq} structure. Let us examine the correctness aspect of the outline. Using the same example and nomenclature: the iterator is on node A, B is masked and before A, C is before B. Since the iterator has node A locked, we know that node B cannot move away. When the \texttt{adl\_node\_force\_pin} call fails, the iterator has to add its waiter to the \texttt{waitq} of B. This must be done before the delete operation has started clearing node B (step (v) of \texttt{adl\_node\_remove\_do}). As the iterator has node A locked, we know the delete operation hasn't even got to the point of updating the pointers to remove B from the list. Hence the step of adding the waiter to the \texttt{waitq} is safe. The iterator will then drop the shared lock on A, thereby stepping out of the way of the delete operation, before calling \texttt{wait}. The iterator still has a pin on node A so that the node will stick around on the list while the iterator waits for the delete of B to finish.

Next, consider the case of insert and delete operations. The problem case of contention in both comes in the \texttt{async\_lock} step where the function tries to take exclusive locks in non-canonical order. The basic operation of taking locks in reverse order under contention was explained in section~\ref{why_async_lock}. Using the nomenclature of that section, let's call the input node as B and the one before it as A. To recap, when the \texttt{async\_lock} call returns FALSE due to contention, the function drops the exclusive lock on B. It then waits for the lock on node A to be granted. Upon receiving the lock on A, the lock on node B is reacquired. This is done in a regular blocking (or synchronous) lock call. 

For correctness, we need to show that the input node B is still valid and that the lock order constraint is not violated. We know that B is still valid as it was pinned upon entering the function and remains pinned (or masked) throughout. Hence, the lock call will not be making an illegal memory access. Initially, node A was before node B and it is now (a) either off the list if the contention was with a delete operation; or (b) one or more nodes have been added between A and B. In case (a), the contending delete operation was stuck as long as the lock on B was held. The \texttt{async\_lock} call put the thread's waiter in line to get the lock on A before stepping out of the delete operations way. Therefore, when the function finally gets the lock on node A, the node is off the list and the contending delete operation is in clear node stage (step (v)) of \texttt{adl\_node\_remove\_do} function. Since A is now off the list, there is no lock order restriction on it with respect to node B (or any other node on the list). This is because no other thread can get to node A anymore. In case (b), new nodes have been added in between but the nodes are still in the canonical order. The input node B is still ahead of node A and reachable by following the chain of \texttt{next} pointers from A. Node A could not have moved ahead of B as that would require deleting node A from the list first. But as we saw in case (a), the delete operation on A would have waited in the clear node stage for this operation and there would be no lock order violation.

From the preceding paragraph, we can see that the \texttt{async\_lock} call will not run into any illegal memory access or deadlock situation. Since the list could have changed, once the lock on node B is reacquired, we do recheck the \texttt{prev} pointer of B to see if it still points to A (which is also locked by the function at this point). If the pointer has changed due to a delete or an insert, then the lock on A is released and the operation is back in its earlier state with only the input node pinned/masked and locked. The operation resumes from that state.
\section{Observations and Extensions}
\label{experience}

We now present some observations on the expected behavior of adlists some of which has also been corroborated by empirical experience of using them. The adlist construct was developed to address the scalability issues of a regular doubly-linked list being seen in the Data Domain File System. It is part of the next major release of the Data Domain OS and has been in active use internally for over a year. It is used in, among other things, simple list based LRU caches, maintaining lists of open connections, open streams and lists of pending IOs.

(i) Since the node delete operation has to wait for the \texttt{pincount} of the node to drop to 0, the thread calling the delete on a node should not hold a pin on any other node. Otherwise, it can get into a circular dependency with another thread and result in a deadlock. The motivation behind splitting the delete node API into 3 steps was to allow a thread to take any action necessary before it blocks to wait for the \texttt{pincount} to reach 0.

(ii) Since any operation on an adlist involves more steps and has to involve more locks than a traditional single mutex approach, replacing a doubly-liked list with an adlist only makes sense for lists seeing a fair amount of contention.

(iii) Since each operation only acquires local locks, each operation is only guaranteed local consistency. For example, an iterator could see the same node twice if another thread removes it from the list after the iterator has crossed it and inserts it back in front of the iterator. Similarly, an iterator can fail to see a node if it moves backward in the list. We have found this to be acceptable in most use cases. For lists that need globally consistent view for certain operations, an adlist needs to be extended with another reader-writer lock. All operations that do not care for a globally consistent view but can modify the list must acquire the reader lock. An operation that requires a globally consistent view must take the writer lock. An iterator that doesn't care about globally consistent view need not worry about the extra lock at all.

(iv) An adlist shows the biggest improvement when the access pattern over the list is scattered or if the list is subject to frequent or long running iterations. If the bulk of the operations are concentrated at any particular location (head or tail for example), then the contention on the lock will reduce the performance to be not much better than a single mutex protected list. There are some remedies in special situations. For example, if a list is being used purely as a queue or a stack, then the lock-free versions of the data structures is a much better choice. For use cases such as an LRU cache, the reclaim operations tend to be limited in parallelism (sometimes only 1 thread can perform it) while re-prioritize operations create pressure on the end of the list (say head) which has ``recent'' nodes. However, since the insert at head for recent nodes does not need to be strictly at the head itself, we can use the following trick: The list is initialized with a certain number of special pre-populated dummy nodes in the head area. An insert at head picks one of the dummy nodes randomly and inserts the node after it. The number of dummy nodes used determines their impact. The count should be high enough to lower contention but not so high that skipping them during iterations becomes an issue. The dummy nodes are also periodically moved back as reclaims would otherwise slowly move them towards the tail.

(v) As is evident from section~\ref{api_contention}, the contention resolution process of adlist requires the ``yielding'' operation to effectively restart. This means that the adlist operations are, in theory, prone to starvation. We have found this to be a non-issue in practice. Although we have some ideas on how to reduce or eliminate this possibility, we haven't pursued them owing to the added complexity and we leave out the details due to space constraints.

\section{Performance}
\label{performance}

We wrote two benchmarks to evaluate the performance of an adlist and its extension that makes use of dummy nodes to lower contention in hot spots of the list. We compare adlists with single mutex protected doubly-linked lists, or ``dlists'' for short from now on. 
The first benchmark simulates a scattered workload assuming uniform access to the lists. The second benchmark simulates the workload typically seen in LRU caches. We present the experimental evaluation for the two benchmarks in sections \ref{benchmark1} and \ref{benchmark2}, respectively.

Each experiment in the benchmarks was run 20 times which was enough to get a confidence level of $99\%$ on the presented average values. All the experiments were carried out on a Intel(R) Xeon(R) Processor E5504. It has 8 cores, each one operating at 2.0GHz and with 4MB of cache. The machine has 12GB of DDR2 memory operating at 1333MHz.

\subsection{Uniform access to the lists}
\label{benchmark1}
In this benchmark each thread is concurrently inserting and removing local elements 
from random points of the list. A thread will only insert and remove its own elements
but it has to iterate over interleaving elements inserted by other threads. 
Each thread inserts and removes batches of $128$ elements and a thousand batches 
are given to each thread. Therefore, each thread inserts and removes $128,000$ elements.
This simulates the behavior of a work-queue where individual requests come and go as they are
received and processed. 

\begin{figure}[htb]
\centering
\includegraphics[scale=0.48]{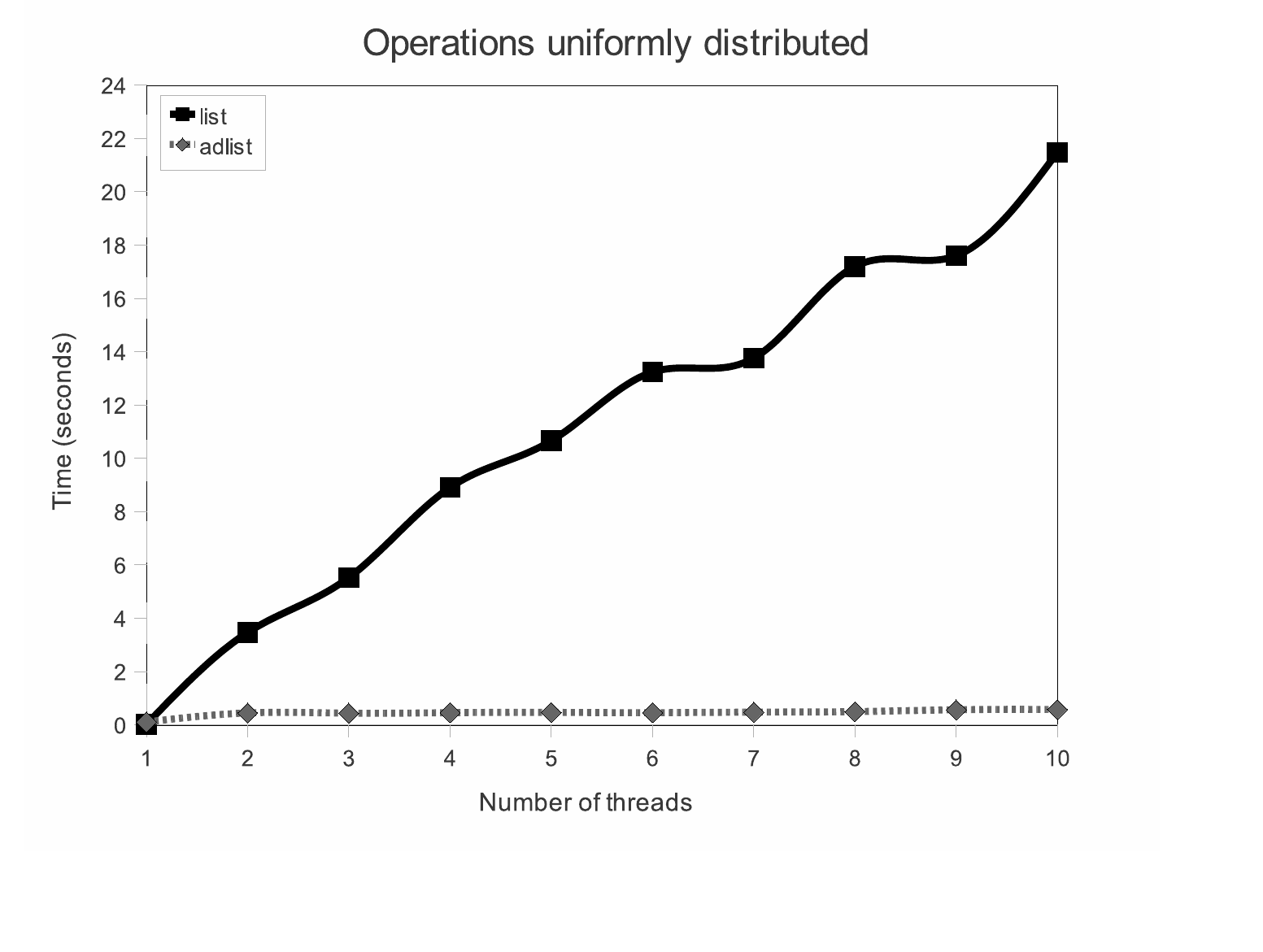}
\caption{Operations uniformly distributed.}
\label{lru_uniform_access}
\end{figure}

Figure~\ref{lru_uniform_access} shows how the total time to run the benchmark varies with the
number of threads. As we throw in more threads the contention increases linearly for dlists whereas
for adlists it stays flat due to the finer-grain locking provided by adlists. For this case the use 
of dummy nodes is not needed because the workload does not create any hot spots in the lists. 

\subsection{LRU workload}
\label{benchmark2}

In this benchmark we mimic workloads seen by LRU caches within the Data Domain File System. There is an array of $\eta$ elements to be inserted so that it is straightforward to have direct access to the elements. Each element is identified by a serial number and has a boolean flag to indicate whether it is present in the LRU cache. All threads carry out the same set of operations and only one of them can evict elements out of the LRU cache at a time. We call the evicting thread ``reclaimer''. To ensure we only measure the contention on the LRU, each thread uses a private lock-free stack to hold elements available to it. This eliminates contention that would otherwise show up from the shared pool of available elements. A thread may execute the following operations:
\begin{enumerate} 
\item [(i)] {\em re-prioritize operation}: an element is removed from its current position in the LRU list and moved to the head of the LRU list (highest priority point).  
\item [(ii)] {\em insert operation}: pop an element from the per-thread lock-free stack of available elements and insert it in at the head of the LRU list. 
\item [(iii)] {\em evict operation}: $k$ elements are removed from the tail of the LRU list and pushed back into the per-thread lock-free stacks in a round-robin fashion. Each element has an eviction cost associated to it which can be set when running the test.
\end{enumerate}
 
\begin{figure}[htb]
\centering
\includegraphics[scale=0.48]{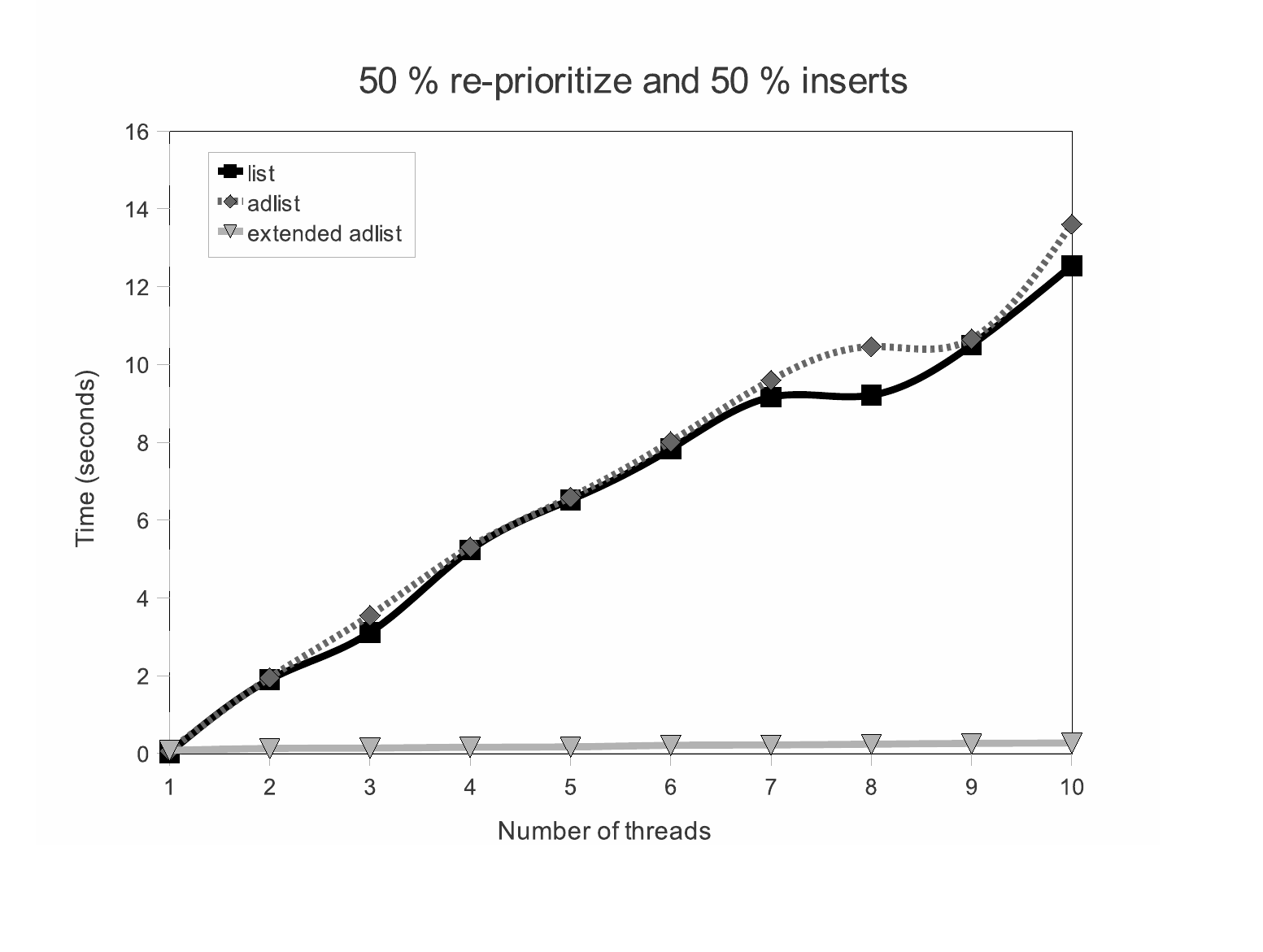}
\caption{Warming up the LRU cache. The extended adlist uses 64 dummy nodes.}
\label{lru_warmup}
\end{figure}

We have looked into three scenarios. The first one is when the cache is being warmed up and therefore each thread tries to find an element to re-prioritize and also inserts a new element. We call this workload ``$50 \%$ re-prioritize and $50 \%$ inserts''. In this workload each thread randomly picks $100,000$ elements, re-prioritizes them if they are found in the LRU cache and also inserts other $100,000$ elements. To have no contention due to eviction, we make $100,000$ elements available per thread. This workload will lead to heavy contention on the head of the LRU list. 
Figure~\ref{lru_warmup} shows how the total time to run the benchmark varies with the
number of threads. 
For this type of workload dlists and adlists behave similarly. However, the extended adlists with dummy nodes leads to great scalability since the total time stays approximately flat as we increase the number of threads and keep the number of operations per thread constant. We fixed the number of dummy nodes at $64$. Later in this section we show how the number of dummy nodes affects the scalability. The main advantage of this extension is that it keeps the simplicity of the code since a single LRU list has to be maintained instead of multiple LRU lists that would be the default extension using dlists.

The second scenario is when eviction happens quite often and its cost dominates the cost of re-prioritize operations. In this workload each thread randomly picks $200,000$ elements, re-prioritizes them if they are found in the LRU cache and also inserts other $20,000$ elements. We call this workload ``$90 \%$ re-prioritize and $10 \%$ inserts''. To create contention due to eviction we make $10,000$ elements available per thread ($50 \%$ of the amount it needs to insert). We also set the eviction cost at $50$ microseconds per element to simulate cases where reclaimation is expensive. The larger this cost, the larger is the contention. When eviction is triggered, the reclaimer thread evicts $k = 100$ elements at a time. This workload leads to contention due to a reclaimer thread iterating from the tail towards the head for a considerable amount of time. Also, most of the operations  over the list are happening on the head which creates a hot spot. 

\begin{figure}[htb]
\centering
\includegraphics[scale=0.48]{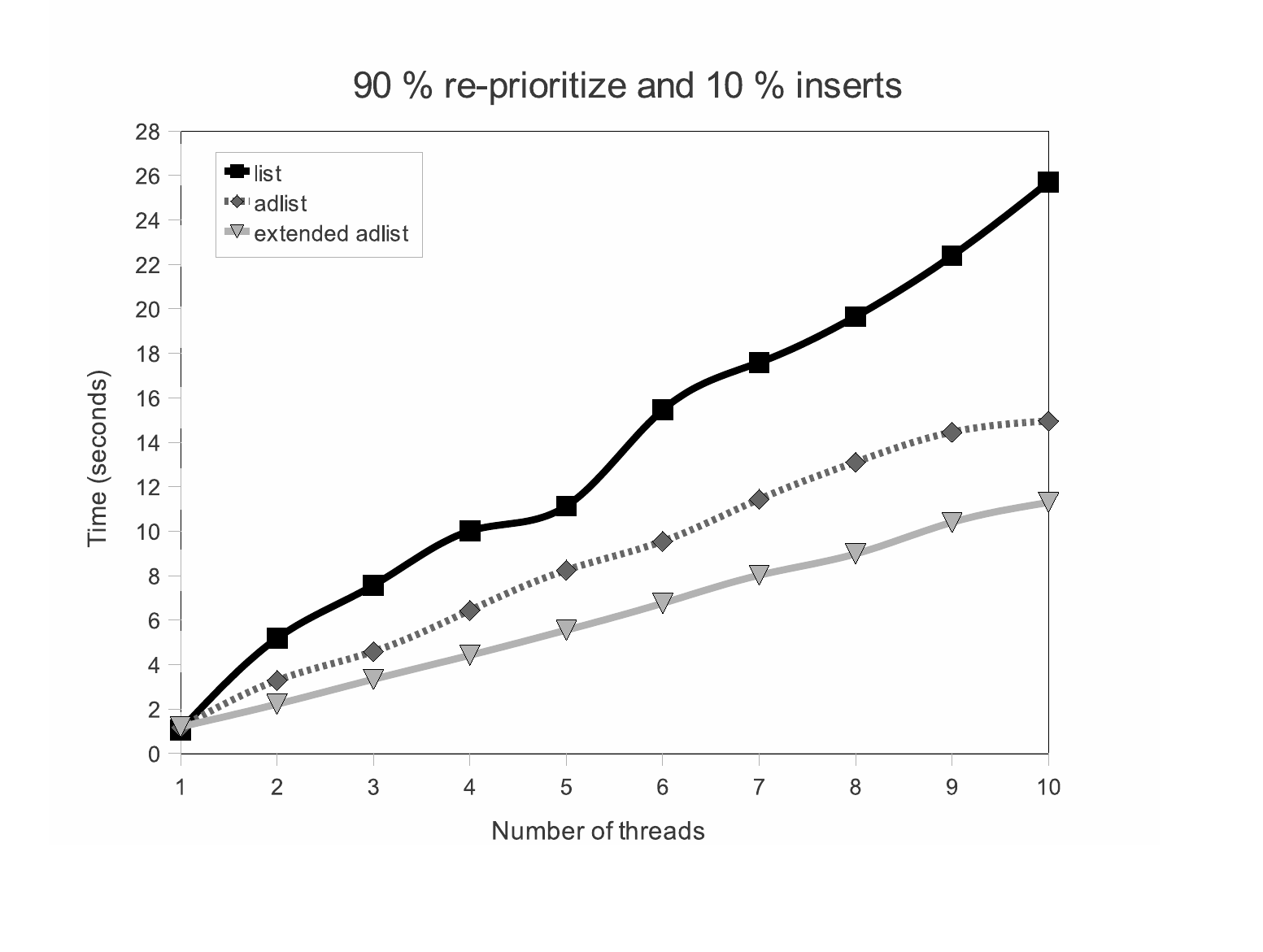}
\caption{Contention due to the cost of evict operations dominating the cost of re-prioritize operations. The extended adlist uses 64 dummy nodes.}
\label{lru_reclaim}
\end{figure}

Figure~\ref{lru_reclaim} shows how the total time to run the benchmark varies with the number of threads. As the number of threads increases we expect the total running time to increase linearly for the three implementations (dlists, adlists and extended adlists). That is because there can be a single reclaimer thread at any point in time. A dlist presents the worse scalability because it can face contention due to all three type of operations: re-prioritize, insert and iteration due to eviction; whereas the adlists do not face contention due to latter. The best scalability comes from an extended adlist with 64 dummy nodes because it only faces contention due to a single reclaimer thread that makes the other threads wait on the availability of elements to insert back.

\begin{figure}[htb]
\centering
\includegraphics[scale=0.48]{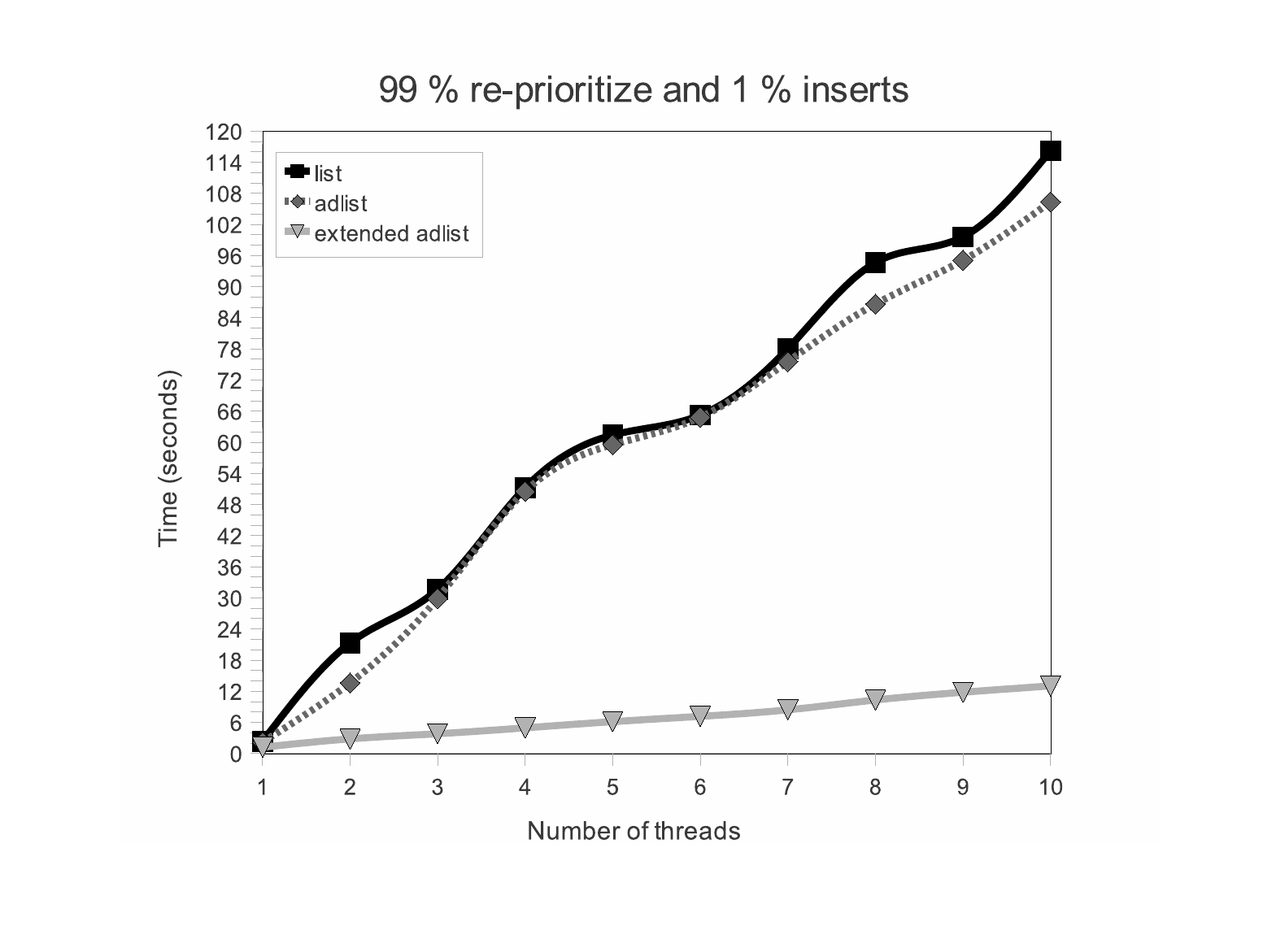}
\caption{Contention due to the cost of re-prioritize operations dominating the cost of evict operations. The extended adlist uses 64 dummy nodes.}
\label{lru_re-prioritize}
\end{figure}

The third scenario is when the cost of re-prioritize operations dominates the cost of evict operations. To achieve that we increase the frequency of re-prioritize operations compared to evict operations. In this workload each thread randomly picks $2,000,000$ elements, re-prioritizes them if they are found in the LRU cache and also inserts other $20,000$ elements. We call this workload ``$99 \%$ re-prioritize and $1 \%$ inserts''. We keep the cost for evict operations the same compared to the previous scenario by using the same afore-described configuration.  As in the first scenario, this workload leads to heavy contention on the head of the LRU list. Therefore dlists and adlists behave similarly whereas extended adlists with 64 dummy nodes scales better. Figure~\ref{lru_re-prioritize} shows how the total time to run the benchmark varies with the number of threads for each of the three implementations.

\begin{figure}[htb]
\centering
\includegraphics[scale=0.48]{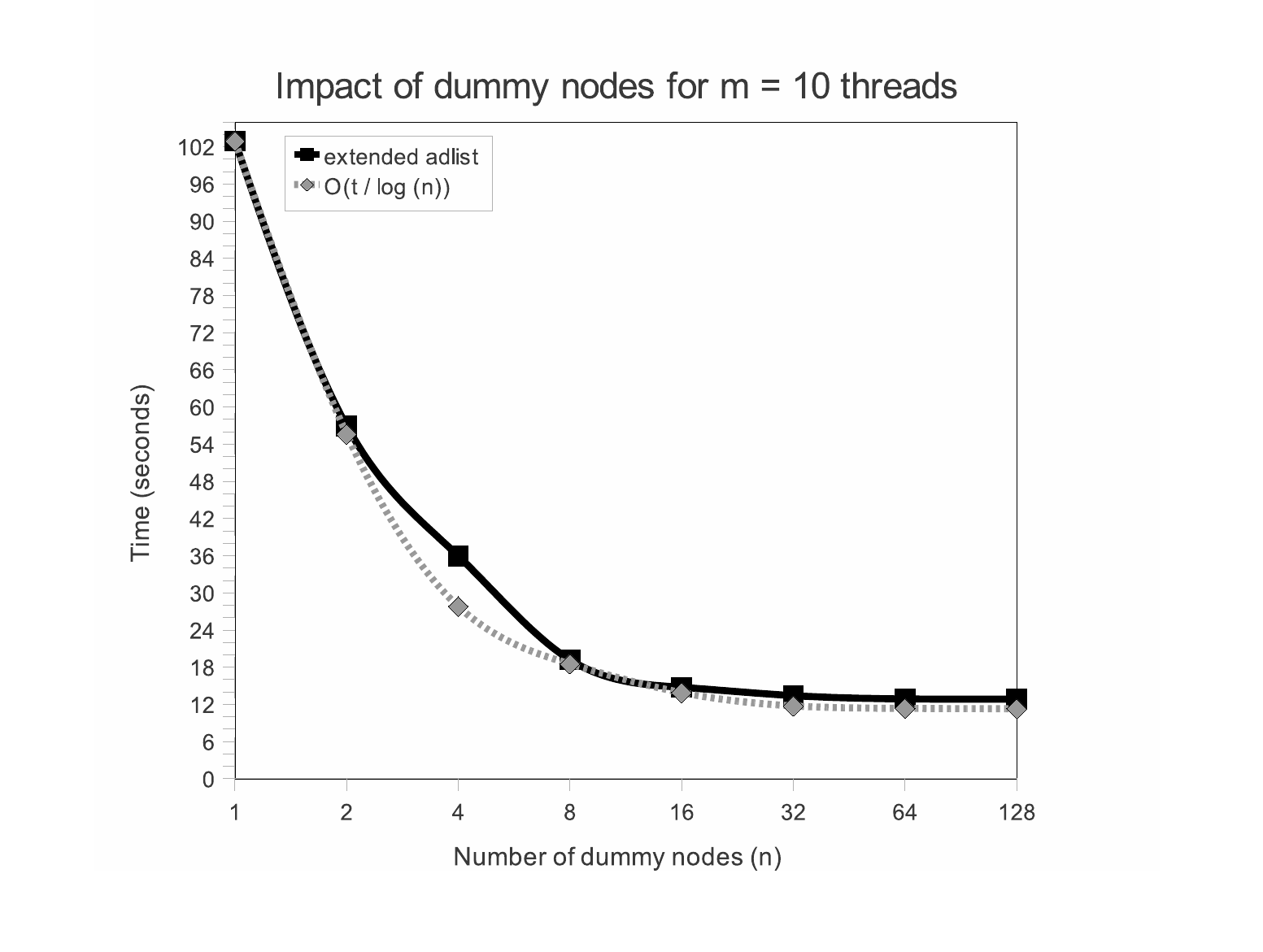}
\caption{Impact of dummy nodes in the runtime with $m = 10$ threads.}
\label{lru_dummy}
\end{figure}

We now evaluate how the dummy nodes affect the runtime for a fixed number of threads. We use the same configuration described above for the scenario three and fix the number of threads in $10$. A single dummy node in an extended adlist makes it equivalent to a normal adlist and all the threads are going to contend on the list's head. To understand the impact of the number of dummy nodes in the total runtime we can see this as the standard ``balls into bins problem'' where $m=10$ threads are thrown into $n$ bins (dummy nodes). The total runtime is inversely proportional to the size of the bin with maximum load (i.e., maximum number of contending threads). A tight analysis for this problem is shown in \cite{rs98} and our experimental data matches their analysis. That is, the runtime is $O(t/log(n))$ where $t$ is the runtime for $n = 1$. Figure \ref{lru_dummy} shows both the experimental and theoretical curves.

\section{Conclusions}
\label{conclusion}

In this paper, we have demonstrated a way of building doubly-linked lists that support greater concurrency than traditional lists. Dubbed adlists, our lists do not place any restriction on the number or memory source of the member nodes. The overhead per node is only 8 bytes which we think should be acceptable in almost all scenarios. The low overhead and high concurrency is made possible due to the use of light-weight synchronization primitives that provide a novel \texttt{async\_lock} mechanism. Using \texttt{async\_lock} allows us to define mechanisms to acquire locks in non-canonical order for a data structure. Being able to do so is crucial to increasing concurrency. Although we have focused exclusively on the list data structures, our approach is general in nature and should be extensible to most other data structures.


\bibliographystyle{plain}

\bibliography{fast2011}

\end{document}